\documentclass[journal,draftcls,onecolumn,12pt,twoside]{IEEEtran}

\usepackage{moreverb}
\usepackage[T1]{fontenc}
\usepackage{hhline}
\usepackage{amssymb,amsfonts,amsthm}
\usepackage{amsmath}
\usepackage{amsfonts}
\usepackage{epsfig}
\usepackage{color}
\usepackage{enumerate}
\usepackage{ifpdf}
\usepackage{graphicx}
\usepackage{cite}
\usepackage{subfig}
\usepackage{caption}
\usepackage{multirow}
\usepackage[table,xcdraw]{xcolor}
\usepackage{mathtools}
\usepackage{algorithm,algpseudocode}
\newcommand\floor[1]{\lfloor#1\rfloor}
\newcommand\ceil[1]{\lceil#1\rceil}

\newtheorem{myprop}{Proposition}

\ifCLASSINFOpdf
\else
\fi

\begin{document}
\title{Optimal Finite Length Coding Rate of Random Linear Network Coding Schemes}
%
%\author{Author 1 and~Author 2	}	% <-this % stops a space
\author{Tan Do-Duy        
        and~M. \'{A}ngeles V\'azquez-Castro,~\IEEEmembership{Senior~Member,~IEEE}% <-this % stops a space
\thanks{The authors are with the Department of Telecommunications and Systems Engineering, School of Engineering, Autonomous University of Barcelona, Spain.
E-mail: \{tan.doduy, angeles.vazquez\}@uab.es} % <-this % stops a space
}

\maketitle
\begin{abstract}
In this paper, we propose a methodology to compute the optimal finite-length coding rate for random linear network coding schemes over a line network. To do so, we first model the encoding, re-encoding, and decoding process of different coding schemes in matrix notation and corresponding error probabilities. Specifically, we model the finite-length performance for random linear capacity-achieving schemes: non-systematic (RLNC) and systematic (SNC) and non-capacity achieving schemes: SNC with packet scheduling (SNC-S) or sliding window (SWNC). Then, we propose a binary searching algorithm to identify optimal coding rate for given target packet loss rate. We use our proposed method to obtain the region of exponential increase of optimal coding rate and corresponding slopes for representative types of traffic and erasure rates. Our results show the tradeoff for capacity-achieving codes vs non-capacity achieving schemes, since the latter trade throughput with delay, which is reflected in the decrease of the exponential slope with the blocklength. We also show the effect of the number of re-encoding times, which further decreases the slope. 
\end{abstract}


%%------------------------------------------------------------------------------------------------------------------------------------------
\section{Introduction \label{Sec:INTRODUCTION}}
Random linear network coding schemes \cite{Shrader.2009, Saxena.2015, Wunderlich.2017, Tan.2015, Li.2011, Li.2012} have emerged as practical network coding (NC) schemes for improving throughput and reliability of wireless networks. Coding operations are executed at the source and at intermediate nodes so that receivers are able to recover from losses at the destination. 

Existing studies on the practical applications of random NC usually assume a fixed size of information source packets \cite{Shrader.2009, Saxena.2015, Wunderlich.2017, Li.2011} or capacity achievability \cite{Pakzad.2005, Saxena.2015} where blocklength, $N$, is assumed to tend to infinity, which is not a practical assumption. However, for the different practical uses of NC, the different design constraints may require different blocklength.
Therefore, in this paper, we study the performance of different random linear NC schemes with finite length coding rate. We consider a simple, yet common in practice, network where a source and a destination are connected over a line network of erasure links e.g., multi-hop device-to-device communications \cite{Pakzad.2005, Nishiyama.2014}. For each NC scheme, we model the encoding, re-encoding, and decoding process in matrix notation and corresponding packet error probability. We then propose a searching algorithm to identify the optimal coding rate for given blocklength and target packet loss rate (PLR).

%%------------------------------------------------------------------------------------------------------------------------------------------
\section{System Model \label{Sec:SYSTEM_MODEL}}
We model a multi-hop line network connecting a source-destination pair as an acyclic graph $\mathcal{G}=(\mathcal{V},\mathcal{E},\mathcal{D})$ where $\mathcal{V}$ and $\mathcal{E}$ are the set of logical associated nodes and directed links in the network, respectively. $\mathcal{D}$ is the set of random variables corresponding to the erasure process associated with each link according to some order \cite{Shi.2013}. We consider random packet erasure model where each link $i$ is modeled as a memoryless channel erased with probability $\delta_{i}$ $(1 \leq i \leq |\mathcal{E}|)$.
We assume a packet stream is produced at the source. NC is applied at all nodes so that some target PLR, $P_e^0$, is guaranteed at the destination.
%%------------------------------------------------------------------------------------------------------------------------------------------
\section{Random Linear Network Coding Schemes \label{Sec:NC_SCHEMES}}
\subsection{Non-Systematic}
For block coding, we consider RLNC and SNC, where information packets are grouped into equally-sized blocks of length $K$ packets. Such a block is also called a generation. For each generation we denote $X_{1\times K}$ as $K$ information packets.

For RLNC, the generator matrix $G\in\mathbb{F}_{q}^{K\times N}$ consists of elements chosen randomly from the same finite field $\mathbb{F}_{q}$.
Let $X'=XG$ represent $N$ coded packets transmitted by the encoder. Coding rate is $\rho=\frac{K}{N}$. 

Each intermediate node re-encodes the linear combinations it has received. The decoding is progressive using Gauss-Jordan elimination algorithm.

The PLR for RLNC over a line path $\mathcal{G}$ is given in Eq. (14) of \cite{Shrader.2009} and written here for convenience
\begin{align}
P_e\left(\mathcal{D},\rho\right) = 1 - \prod_{i=1}^{|\mathcal{E}|} \sum_{n=K}^{N} \prod_{j=0}^{K-1}(1-q^{j-n})Pr(N_i=n),
\label{eq:PE_RLNC}
\end{align}
where $N_{i}\sim bin(N,1-\delta_i)$ are binomial random variables denoting number of coded packets received at re-encoding/decoding node $i$ $(1\le i \le |\mathcal{E}|)$.
\subsection{Systematic}
The SNC generator matrix is $G=\left[\begin{matrix} I_{K} ~ C \end{matrix}\right]$. It consists of the identity matrix $I_{K}$ of dimension $K$ and a coefficient matrix $C\in\mathbb{F}_{q}^{K \times (N-K)}$ with elements randomly chosen from the same finite field $\mathbb{F}_{q}$.
Let $X'=XG$ represent $K$ information (systematic) packets and $N-K$ coded packets transmitted by the encoder. Coding rate is $\rho=\frac{K}{N}$.

If a systematic packet is lost, the re-encoder sends a random linear combination of the information packets stored in its buffer \cite{Saxena.2015}.
The decoding is progressive using Gauss-Jordan elimination algorithm.

The PLR for SNC over a line path $\mathcal{G}$ is given in Eqs. (15)-(18) of \cite{Shrader.2009} and written here for convenience
\begin{align}
P_e\left(\mathcal{D},\rho\right) = 1 - \frac{1}{K} \sum_{z=1}^{K}zPr(Z_{|\mathcal{E}|} = z),
\label{eq:PE_SNC}
\end{align}
where $Z_{|\mathcal{E}|}$ is the random number of packets decoded at the decoder.
%with $Z_{|\mathcal{E}|}$, random number of packets decoded at the decoder.
%
\subsection{Systematic with scheduling}
\begin{figure*}[htbp]
\begin{centering}
\subfloat[SNC							\label{fig:ILLUSTRATION_SRNC}]				{\epsfig{file=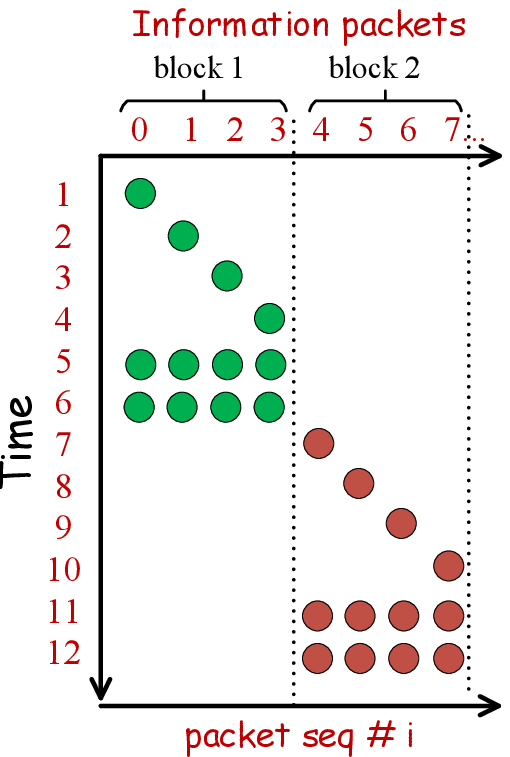,width=0.30\textwidth}}
\hspace{2 mm}
\subfloat[SNC-S						\label{fig:ILLUSTRATION_PACE}]				{\epsfig{file=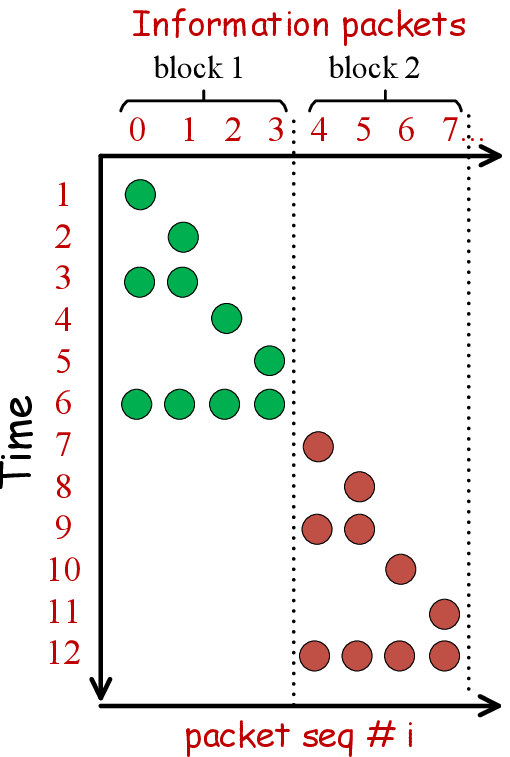,width=0.30\textwidth}}
\hspace{2 mm}
\subfloat[SWNC						\label{fig:ILLUSTRATION_SWNC}]				{\epsfig{file=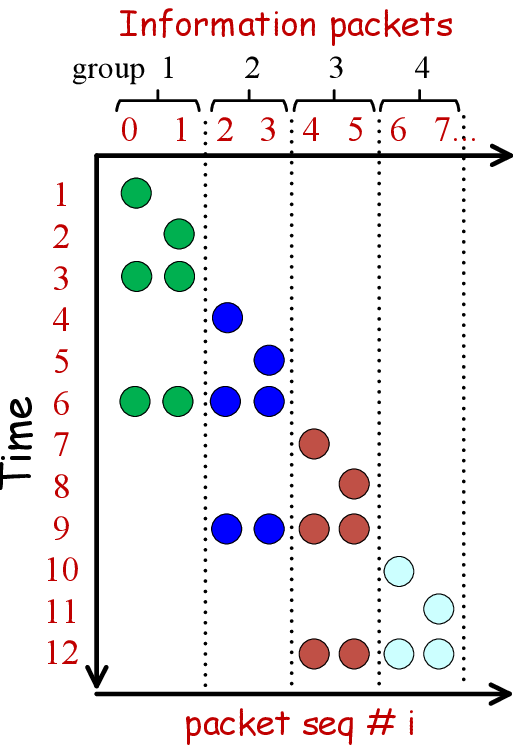,width=0.314\textwidth}}
\par\end{centering}
\caption{Illustration of SNC, SNC-S with $K=4$, SWNC with $w_e=4$ for $\rho=\frac{2}{3}$. Each single dot represents a uncoded packet whereas multiple dots on a line represent a coded packet.
\label{fig:ILLUSTRATION_SRNC_SWNC}}
\end{figure*}

The systematic generator $G=\left[\begin{matrix} I_{K} ~ C \end{matrix}\right]$ can be modified to allow variable scheduling of packets (a similar approach is in \cite{Wunderlich.2017}). Here, we consider the case illustrated in Fig. \ref{fig:ILLUSTRATION_PACE}, the first $\frac{K}{2}$ information packets of each generation are transmitted first and followed by $n_c$ coded packets which are created by random linear combinations of the first $\frac{K}{2}$ information packets. The last $\frac{K}{2}$ information packets and $n_c$ coded packets are transmitted in sequence. The last $n_c$ coded packets combine all $K$ information packets.
In matrix notation, we model SNC with scheduling (SNC-S) as follows. Let $X' = XG$ represent information packets and $2n_c$ coded packets sent by the encoder during $N=K+2n_c$ timeslots. Coding rate is $\rho=\frac{K}{N}$. 

The generator matrix is $G = \big[I_1 ~ C_1 ~ I_2 ~ C_2\big]$, where 
\begin{itemize}
	\item $I_1$ : the matrix of size $K\times \frac{K}{2}$, the first $\frac{K}{2}$ columns of the identity matrix of dimension $K$.
	\item $C_1$: the matrix of size $K\times n_c$ where elements of the first $\frac{K}{2}$ rows are chosen randomly from the same $\mathbb{F}_{q}$ while elements of the last $\frac{K}{2}$ rows are zero.
	\item $I_2$ : the matrix of size $K\times \frac{K}{2}$, the last $\frac{K}{2}$ columns of the identity matrix of dimension $K$.	
	\item $C_2$: the matrix of size $K\times n_c$ with all elements chosen randomly from the same $\mathbb{F}_{q}$.
\end{itemize}

The re-encoder immediately forwards any received information packet to the next node. If a systematic packet is lost, the re-encoder sends a random linear combination of all packets according to the same block stored in its buffer. 
The decoding is progressive using Gauss-Jordan elimination algorithm.

Due to the lack of analytical expressions, we use simulation results to identify the PLR $P_e(\mathcal{D},\rho)$ for SNC-S.
\subsection{Sliding Window}
We also consider a finite sliding window RLNC scheme (SWNC) \cite{Wunderlich.2017} (illustrated in Fig. \ref{fig:ILLUSTRATION_SWNC}). The encoder uses an encoding window of size $w_e$ successive information packets.
Specifically, the sequence of information packets is divided into groups of size $K_s=\frac{w_e}{2}$. When a new $x$-th group is generated, the encoding window slides over the sequence of information packets by appending $K_{s}$ packets of the new group to the end of the window and removing the $(x-2)$-th group from the beginning of the window $(x\geq 3)$. $K_{s}$ information packets of the new group are sent in consecutive timeslots followed by $n_{c}$ coded packets created by random combinations of $w_e$ information packets in the encoding window. Coding rate is $\rho=\frac{w_{e}}{N}$ with $N=w_{e}+2n_{c}$.

The generator matrix $G\in\mathbb{F}_{q}^{w_e \times n_{c}}$ is a coding coefficient matrix with each element $g_i^j$ denoting a coefficient corresponding to an information packet $i$-th to obtain the $j$-th coded packet $(i,j=0,1,...)$.
We denote $X'=XG$ as $n_{c}$ coded packets with $X_{1\times w_e}$ denoting $w_e$ information packets covered by the encoding window.

%% RE-ENCODING
If an information packet with sequence number $i$ is lost, the re-encoder sends a random combination of information packets stored in its buffer with sequence number in range $\big[0, \frac{w_e}{2}-1\big]$ if $x=1$ or $\big[(x-2)\frac{w_e}{2}, x\frac{w_e}{2}-1\big]$ if $x \geq 2$ with $x=\lfloor \frac{i}{w_e/2} \rfloor + 1$.

We denote $w_d$, $w_d \geq w_e$, as the finite decoding window size in number of successive packets e.g. $w_d=w_e$. Iterative Gaussian Elimination can be used to decode packet by packet \cite{Wunderlich.2017}. 

Due to the lack of analytical expressions, we use simulation results to identify the PLR $P_e(\mathcal{D},\rho)$ for SWNC.

\begin{figure*}[htbp]
\begin{centering}
\subfloat[$\delta_i=0.05$				\label{fig:PLR_vs_CodeRate_E005}]			{\epsfig{file=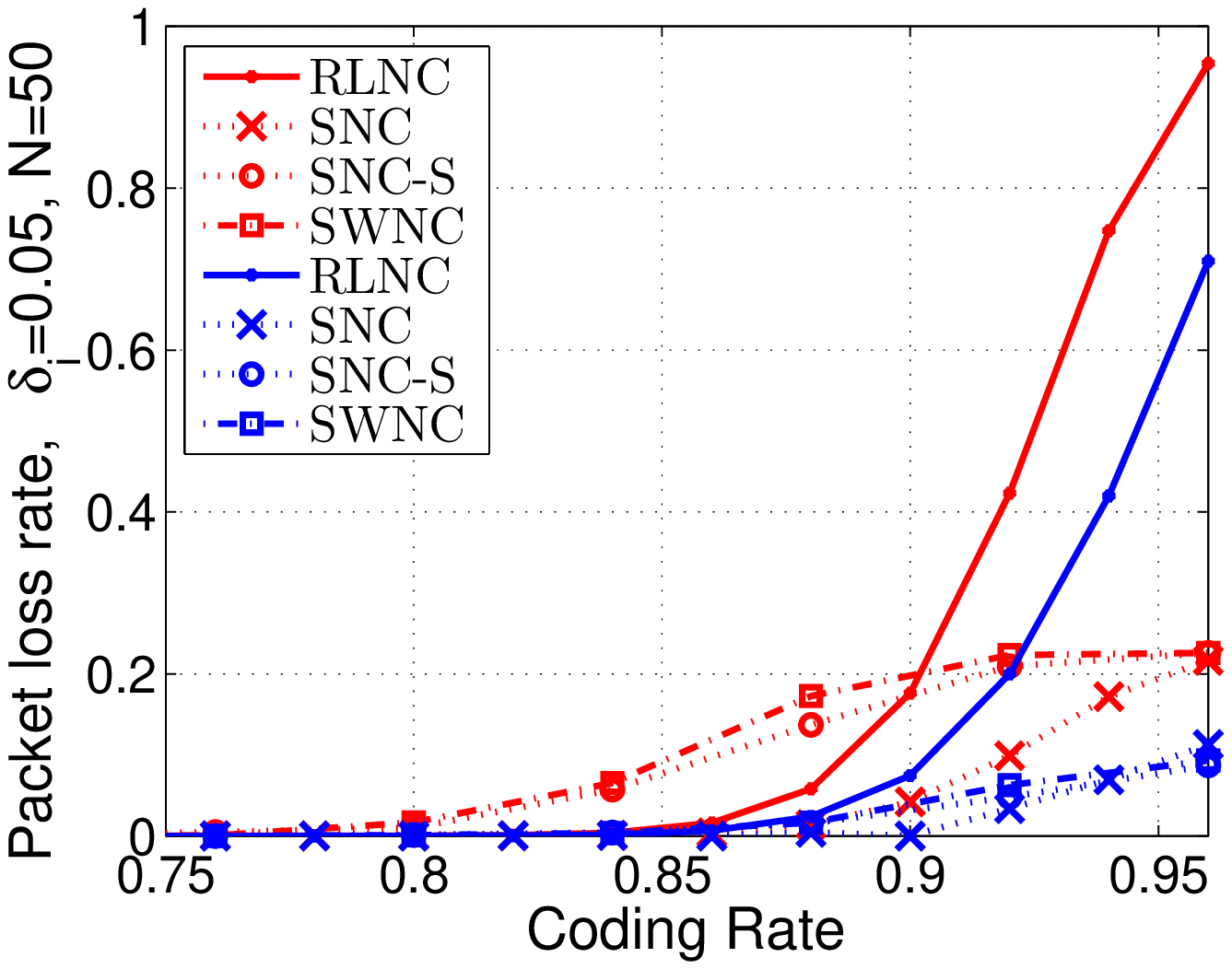,width=0.45\textwidth}}
\hspace{10mm}
\subfloat[$\delta_i=0.2$				\label{fig:PLR_vs_CodeRate_E02}]			{\epsfig{file=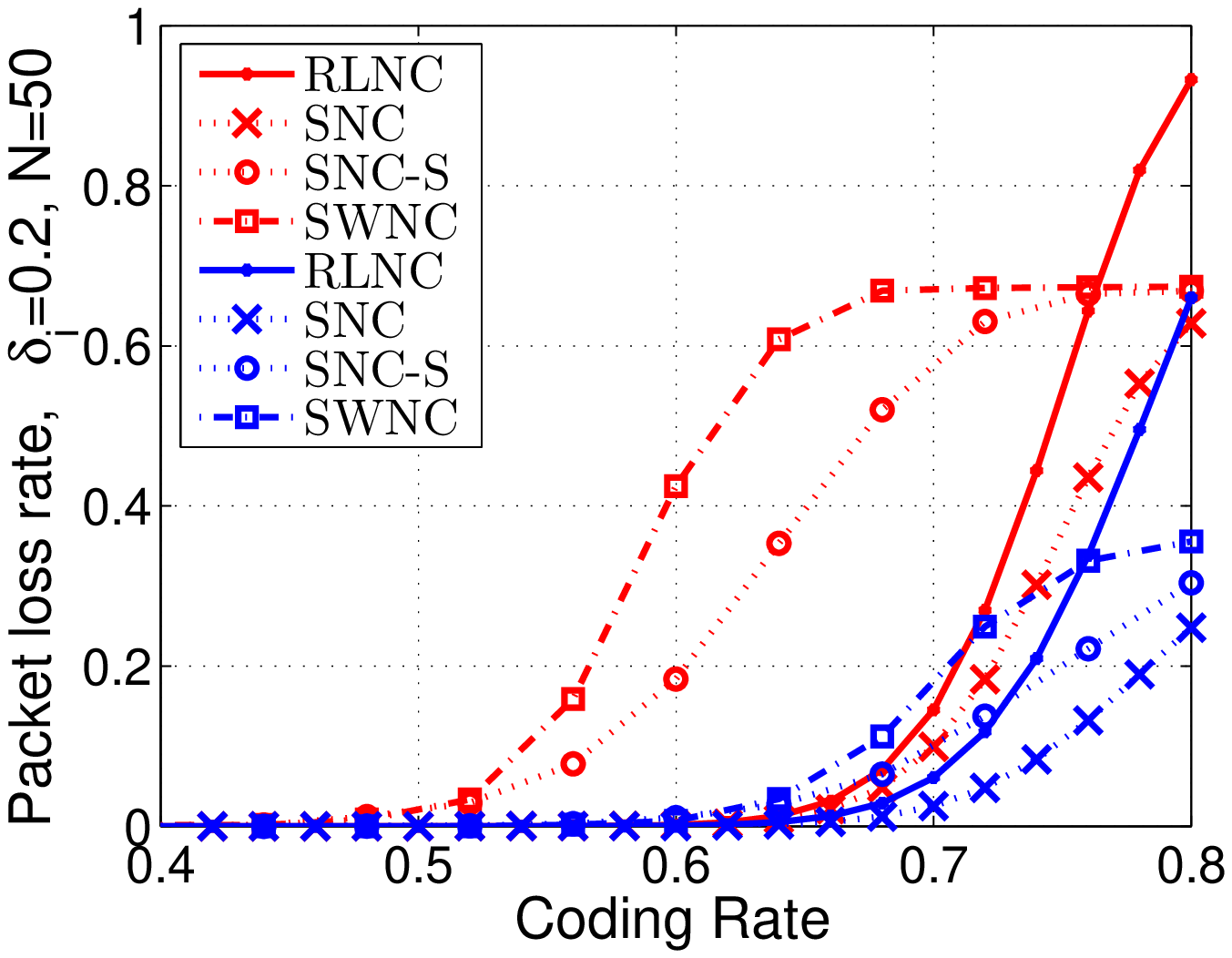, width=0.45\textwidth}}
\par\end{centering}
\caption{PLR, $P_e(\mathcal{D},\rho)$, over 2-hop (blue) and 5-hop (red) line networks with $\delta_i=0.05$ and $0.2$ $(\forall i \in [1,|\mathcal{E}|])$. PLR for RLNC and SNC is calculated using Eqs. (\ref{eq:PE_RLNC})-(\ref{eq:PE_SNC}) while PLR for SNC-S and SWNC is averaged through simulations.
\label{fig:PLR_vs_CodeRate}}
\end{figure*}

For illustration, in Fig. \ref{fig:PLR_vs_CodeRate}, we show numerical values of PLR as a function of $\rho$ for the considered NC schemes on 2-hop and 5-hop line networks.
%
%%------------------------------------------------------------------------------------------------------------------------------------------
\section{Optimal Finite Length Coding Rate \label{Sec:DERIVATION}}
We now investigate how to choose an optimal coding rate, $\rho^*$, given blocklength $N$ for some target $P_e^{0}$ at the destination node in $\mathcal{G}$ with $|\mathcal{E}|$ links (hops).

\subsection{Proposed searching algorithm}
We show that a searching algorithm is sufficient to identify $\rho^*$ given $N$ and $P_e^{0}$ for random linear NC schemes.
\begin{myprop} \label{PROP: PLR_BOUND_RLNC}
PLR for RLNC can be bounded as
\begin{equation}
1-\prod_{i=1}^{|\mathcal{E}|}1-C_{p_i} \big(K-1\big) \leq P_e\left(\mathcal{D},\rho\right) \leq 1-\prod_{i=1}^{|\mathcal{E}|}1-C_{p_i}\big(K\big),	%\nonumber
\label{eq:PE_RLNC3}
\end{equation}
where $C_{p_i}(m)=\phi\Big(sgn(\frac{m}{N}-p_i)\sqrt{2NH(\frac{m}{N},p_i)}\Big)$, with $sgn(x)$ be the usual signum function with argument $x$, $\phi(y)$ be the distribution function of a standard normal variable with argument $y$, $H(x,p_i )=xln\frac{x}{p_i} +(1-x)ln\frac{1-x}{1-p_i}$, $sgn(x)=\frac{x}{|x|}$, $p_i=1-\delta_i$.
\end{myprop}

\begin{IEEEproof} 
For a sufficiently large field size $q$, PLR for RLNC given in Eq. (\ref{eq:PE_RLNC}) can be rewritten as
$P_e\left(\mathcal{D},\rho\right) = 1 - \prod_{i=1}^{|\mathcal{E}|} \sum_{n=K}^{N} Pr(N_i=n)$. 
Zubkov et al. \cite{Zubkov.2013} proved that for $1\leq m \leq N$, $C_{p_i}(m)\leq \sum_{n=0}^{m} Pr(N_i=n)\leq C_{p_i}(m+1)$. Hence, we obtain the bound of PLR for RLNC as in Eq. (\ref{eq:PE_RLNC3}).
\end{IEEEproof} 

\begin{myprop} \label{PROP: SEARCHING_ALGORITHM}
A searching algorithm is sufficient to identify the optimal coding rate $\rho^*$ for some given $N$ and $P_e^{0}$.
\end{myprop}
\begin{IEEEproof} 
We consider PLR for RLNC at the upper bound proposed in Proposition \ref{PROP: PLR_BOUND_RLNC}, given as
\begin{align}
P_e\left(\mathcal{D},\rho\right) = 1-\prod_{i=1}^{|\mathcal{E}|}1-\phi\Big(sgn(\rho-p_i)\sqrt{2NH(\rho,p_i)}\Big).
\label{eq:PE_RLNC4}
\end{align}
Let consider functions $z= H(\rho,p_i)$, $y=sgn(\rho-p_i)\sqrt{2NH(\rho,p_i)}$, and $f(\rho)=1-\prod_{i=1}^{|\mathcal{E}|}1-\phi\Big(y(\rho)\Big)$.
We obtain $z^{'}=\frac{dz}{d\rho}=ln \frac{\rho(1-p_i)}{(1-\rho)p_i}=0$ at $\rho=p_i$. 
For example, for $p_i=0.8,\rho=0.5$, $z^{'}=ln(0.25)<0$. Hence, $z$ decreases with $\rho\in (0,p_i)$ and then increases with $\rho\in (p_i,1)$.
Further, for $\rho<p_i$, $sgn(\rho-p_i )=-1$. Hence, $y$ increases with $0<\rho<1$. Thus, $f(\rho)$ is a non-decreasing function of $\rho$. For some $N$, $f(\frac{K}{N})$ is also a non-decreasing function of $K$.
Hence, assume $K=N\rho$, a unique $\rho^*$ can be identified by searching algorithms \cite{Dalal.2004} so that $P_e\left(\mathcal{D},\rho^*\right) \simeq P_e^0$.
\end{IEEEproof} 

\begin{algorithm}
\caption{The proposed searching algorithm to identify $\rho^*$.}
\label{Algo:SEARCHING}
\begin{algorithmic}[1]
\State 			\textbf{Initialize}
\State \hspace{8pt}	Set $\mathcal{D}$, $\mathcal{E}$, $N$, $P_e^0$, $K=\ceil{N\rho}$ (or $w_e=\ceil{N\rho}$ in SWNC)
\State \hspace{8pt}	For a set $\Psi$ of coding rate, $\rho[i]\in \Psi$, $1\leq i\leq |\Psi|$  
\State \hspace{8pt}	Set $first=1$, $last=|\Psi|$, $\rho^*=\emptyset$
\State 			\textbf{while}	$(first<last)$	\hspace{4mm}	\{
\State \hspace{10pt}		Set $mid=\floor{\frac{first+last}{2}}$
\State \hspace{10pt}		Calculate $P_e(\mathcal{D},\rho[mid])$	using theoretical expressions or simulation
\State \hspace{10pt}  \textbf{if}			 $mid=first$	\hspace{2mm}	\textbf{then}		
\State \hspace{30pt}							\textbf{if} 	$P_e(\mathcal{D},\rho[mid])\le P_e^0$			\hspace{2mm}	\textbf{then}			
\State \hspace{45pt}																	Set $\rho^{*}=\rho[mid]$
\State \hspace{30pt}							\textbf{break}																																			
\State \hspace{10pt}	\textbf{else~if} $P_e(\mathcal{D},\rho[mid])\le P_e^0$			\hspace{2mm}				\textbf{then}
\State \hspace{30pt}							Set $first=mid$, $\rho^{*}=\rho[mid]$
\State \hspace{10pt}  \textbf{else}		\hspace{1mm}		Set $last=mid$	\hspace{2mm}	\}	\hspace{10pt}		%$P_e(\mathcal{D},\rho[mid]) > P_e^0$		
\State    	Return $\rho^*$
\end{algorithmic}
\end{algorithm}

Hence, we propose an algorithm based on the binary searching algorithm in \cite{Dalal.2004}, which is summarized in Algorithm \ref{Algo:SEARCHING}.
By assuming $K=\ceil{N\rho}$ (or $w_e=\ceil{N\rho}$), PLR for a specific NC scheme is considered as a function of $\rho$ and $|\mathcal{E}|$. $\rho^*$ is the maximum $\rho$ that $P_e(\mathcal{D},\rho) \leq P_e^0$ holds. We denote $\Psi$ as the set of coding rates available for searching. Then, the complexity of the proposed searching algorithm is $\mathcal{O}(log_2|\Psi|)$ \cite{Dalal.2004}.

\subsection{Performance evaluation}
We evaluate the performance of random linear NC schemes with $P_e^0=10^{-6}$ and $P_e^0=10^{-3}$ representing the quasi-error-free channel and delay-constrained applications \cite{Chen.2004}, respectively. In each case, we consider $\delta_i=0.05$ and $\delta_i=0.2$ $(\forall i \in [1,|\mathcal{E}|])$ representing 802.11 wireless links \cite{Salyers.2008} and transmission scenarios with links having light rainfall \cite{Saxena.2015,Cacheda.2007}, respectively. However, our proposed Algorithm \ref{Algo:SEARCHING} can operate with any target PLR and erasure rates.

\begin{figure*}[htbp]
\begin{centering}
\subfloat[$\delta_i=0.05$				\label{fig:CODERATE_E005_Pe10e6}]			{\epsfig{file=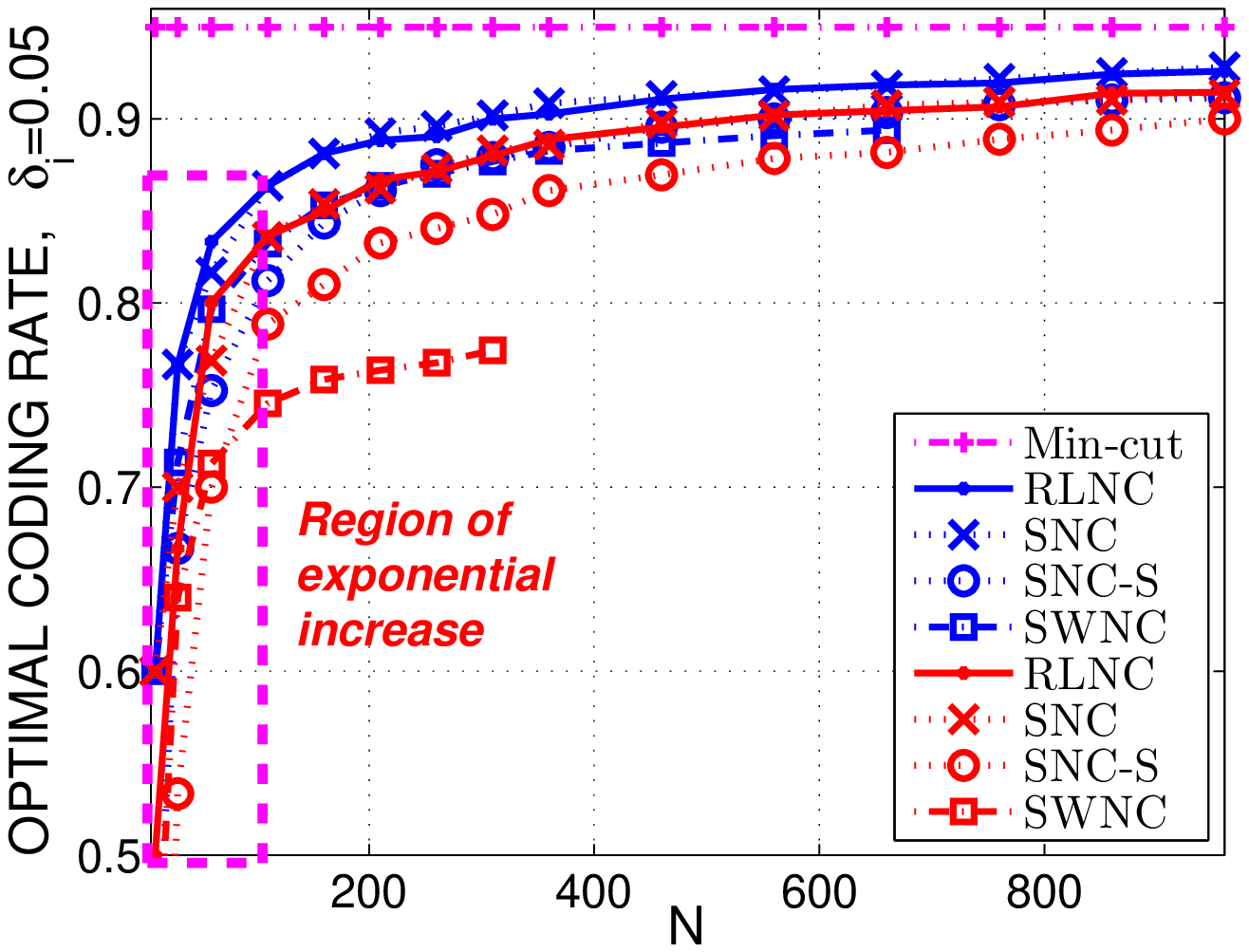,width=0.45\textwidth}}
\hspace{10mm}
\subfloat[$\delta_i=0.2$				\label{fig:CODERATE_E02_Pe10e6}]			{\epsfig{file=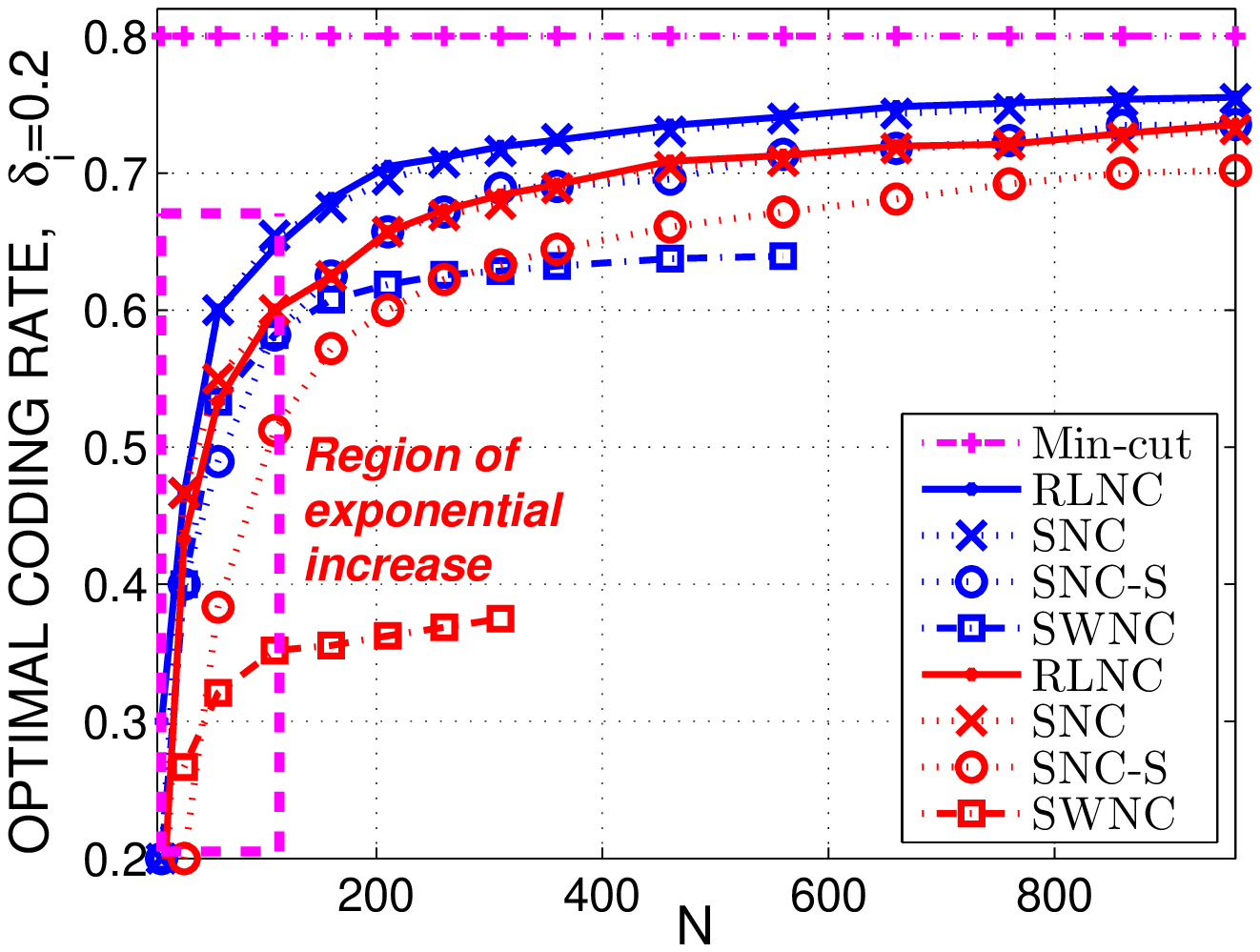,width=0.45\textwidth}}
\par\end{centering}
\caption{Optimal coding rate, $\rho^*$, versus $N$ for different NC schemes over 2-hop and 5-hop line networks with $\delta_i=0.05$ and $\delta_i=0.2$ $(\forall i \in [1,|\mathcal{E}|])$, respectively, for $P_e^0=10^{-6}$ (blue: $|\mathcal{E}|=2$ hops, red: $|\mathcal{E}|=5$ hops). For RLNC and SNC, $\rho^*$ is calculated using Eqs. (\ref{eq:PE_RLNC})-(\ref{eq:PE_SNC}) whereas for SNC-S and SWNC, $\rho^*$ is averaged through multiple simulations.
\label{fig:OPTIMAL_CODERATE}}
\end{figure*}

\begin{figure*}[htbp]
\begin{centering}
\subfloat[$\delta_i=0.05$				\label{fig:CODERATE_E005_Pe10e6_N100}]		{\epsfig{file=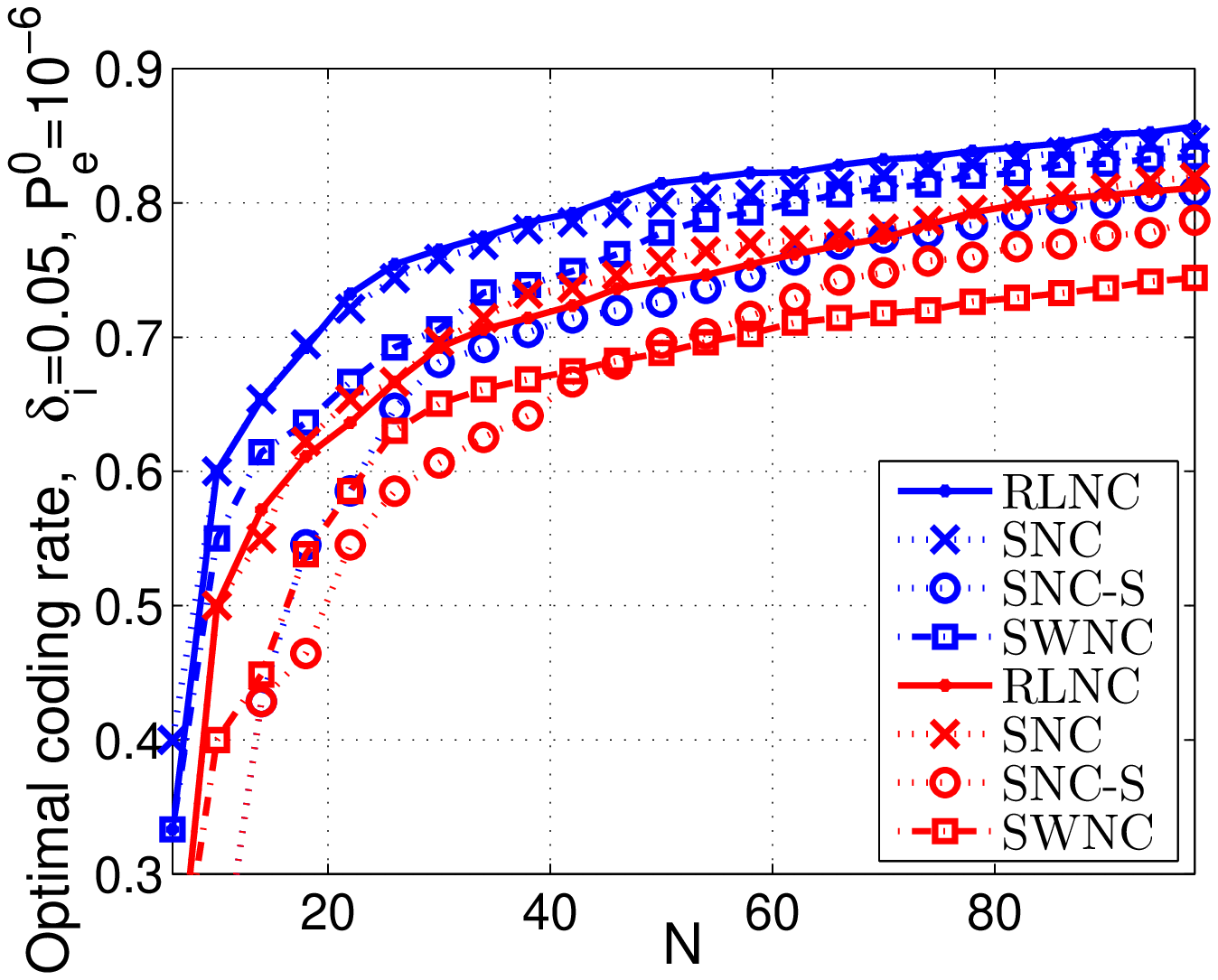,width=0.45\textwidth}}
\hspace{10mm}
\subfloat[$\delta_i=0.2$				\label{fig:CODERATE_E02_Pe10e6_N100}]			{\epsfig{file=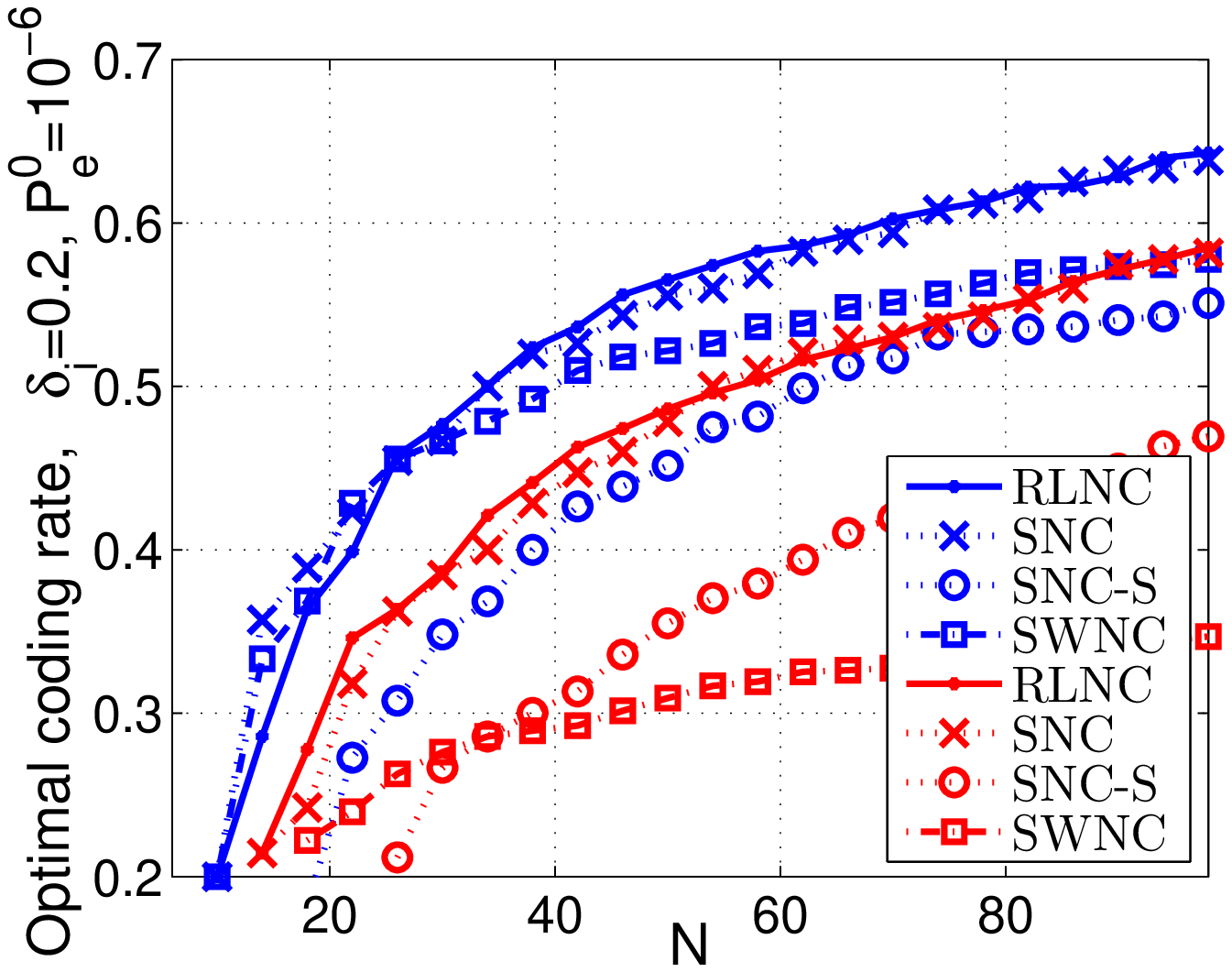,width=0.45\textwidth}}
\hfill
\subfloat[$\delta_i=0.05$				\label{fig:CODERATE_E005_Pe10e3_N100}]			{\epsfig{file=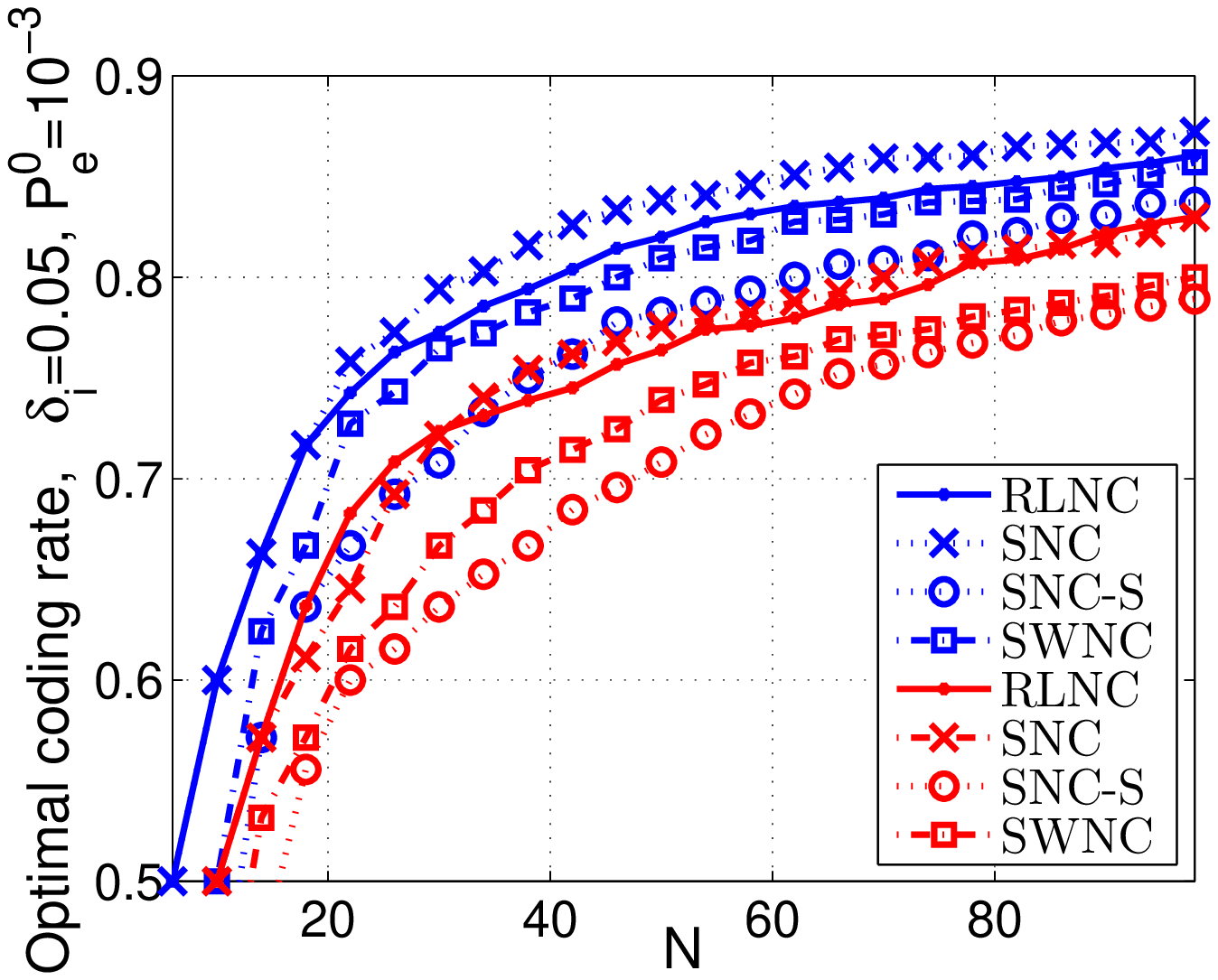,width=0.45\textwidth}}
\hspace{10mm}
\subfloat[$\delta_i=0.2$				\label{fig:CODERATE_E02_Pe10e3_N100}]				{\epsfig{file=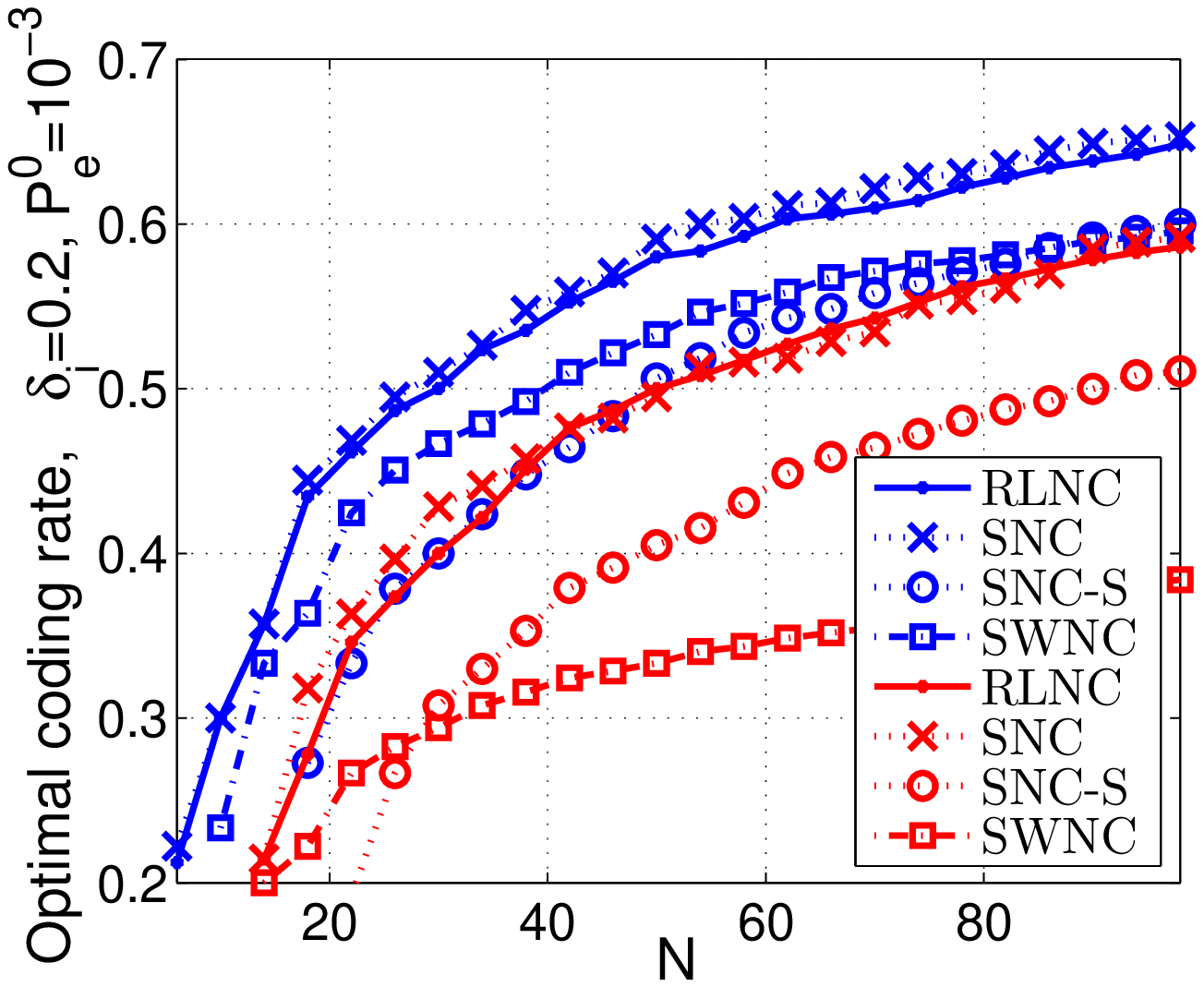,width=0.45\textwidth}}
\par\end{centering}
\caption{Optimal coding rate in the region of exponential increase $(N \le 100)$ with $\delta_i=0.05$ and $\delta_i=0.2$ for $P_e^0=10^{-6}$ (Fig. \ref{fig:CODERATE_E005_Pe10e6_N100}-\ref{fig:CODERATE_E02_Pe10e6_N100}) and $P_e^0=10^{-3}$ (Fig. \ref{fig:CODERATE_E005_Pe10e3_N100}-\ref{fig:CODERATE_E02_Pe10e3_N100}) (blue: $|\mathcal{E}|=2$ hops, red: $|\mathcal{E}|=5$ hops).
\label{fig:OPTIMAL_CODERATE_Pe0_N100}}
\end{figure*}

\begin{figure*}[htbp]
\begin{centering}
\subfloat[$\delta_i=0.05$				\label{fig:SLOPE_E005_Pe10e6_N100}]			{\epsfig{file=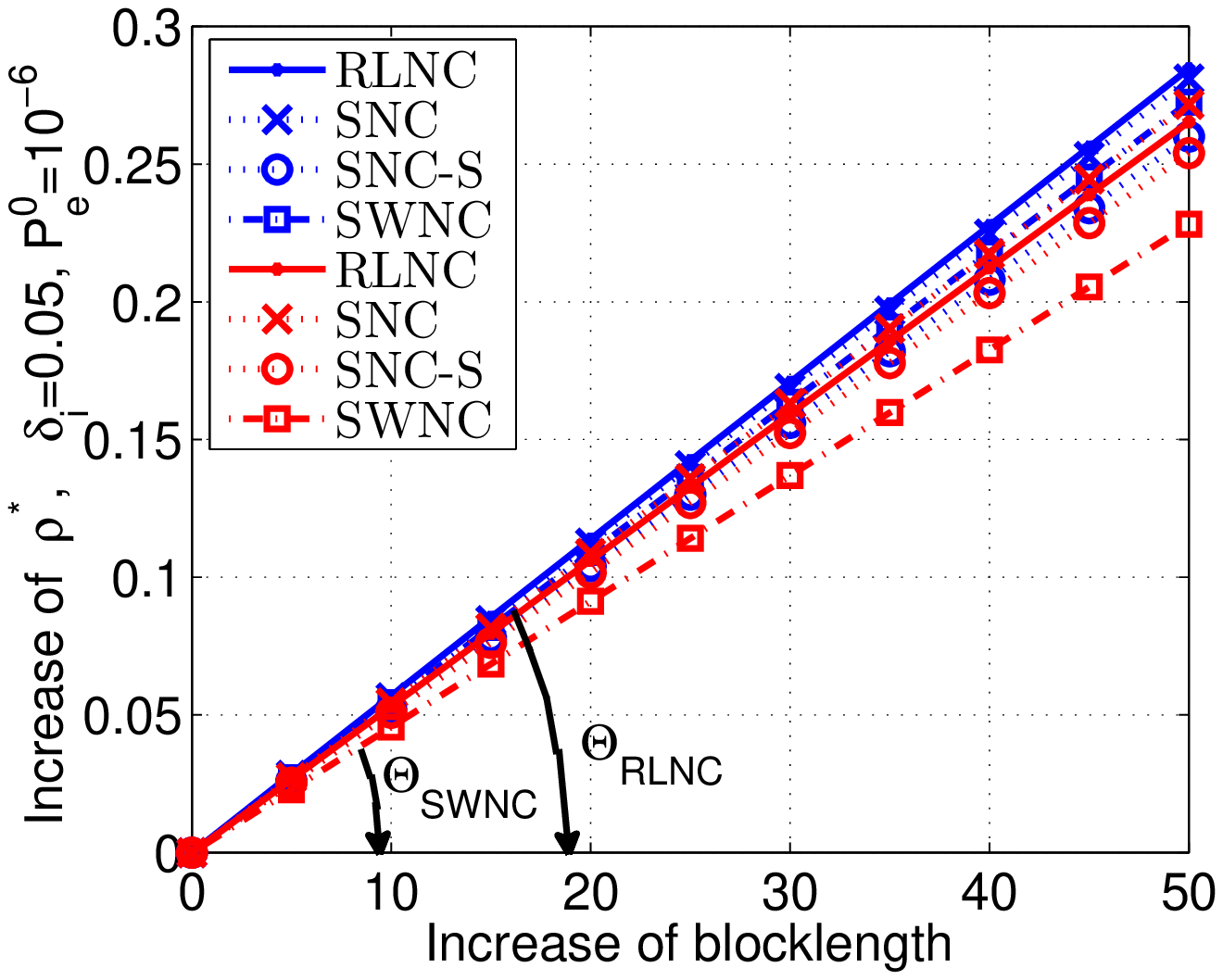,width=0.45\textwidth}}
\hspace{10mm}
\subfloat[$\delta_i=0.2$				\label{fig:SLOPE_E02_Pe10e6_N100}]			{\epsfig{file=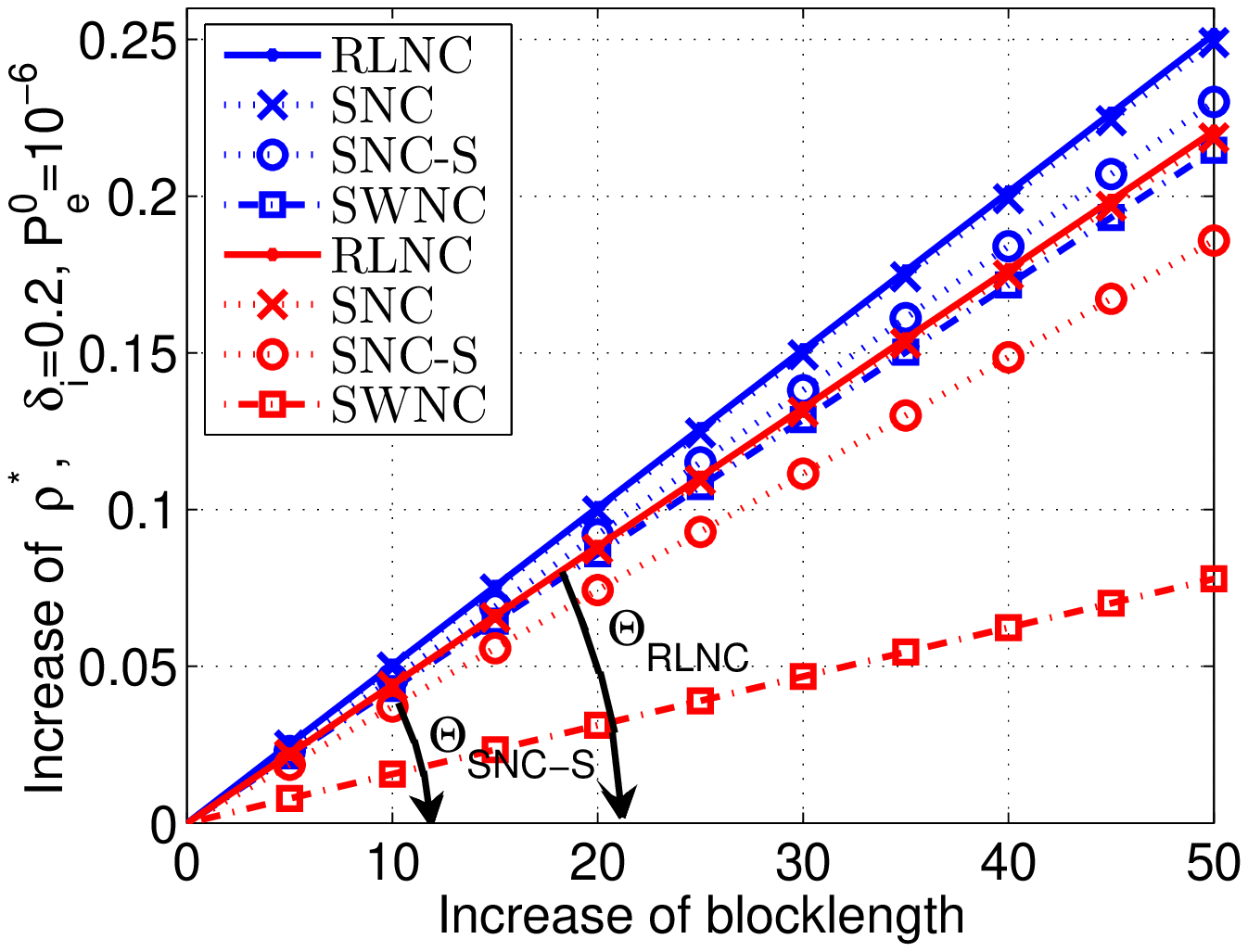,width=0.45\textwidth}}
\hfill
\subfloat[$\delta_i=0.05$				\label{fig:SLOPE_E005_Pe10e3_N100}]			{\epsfig{file=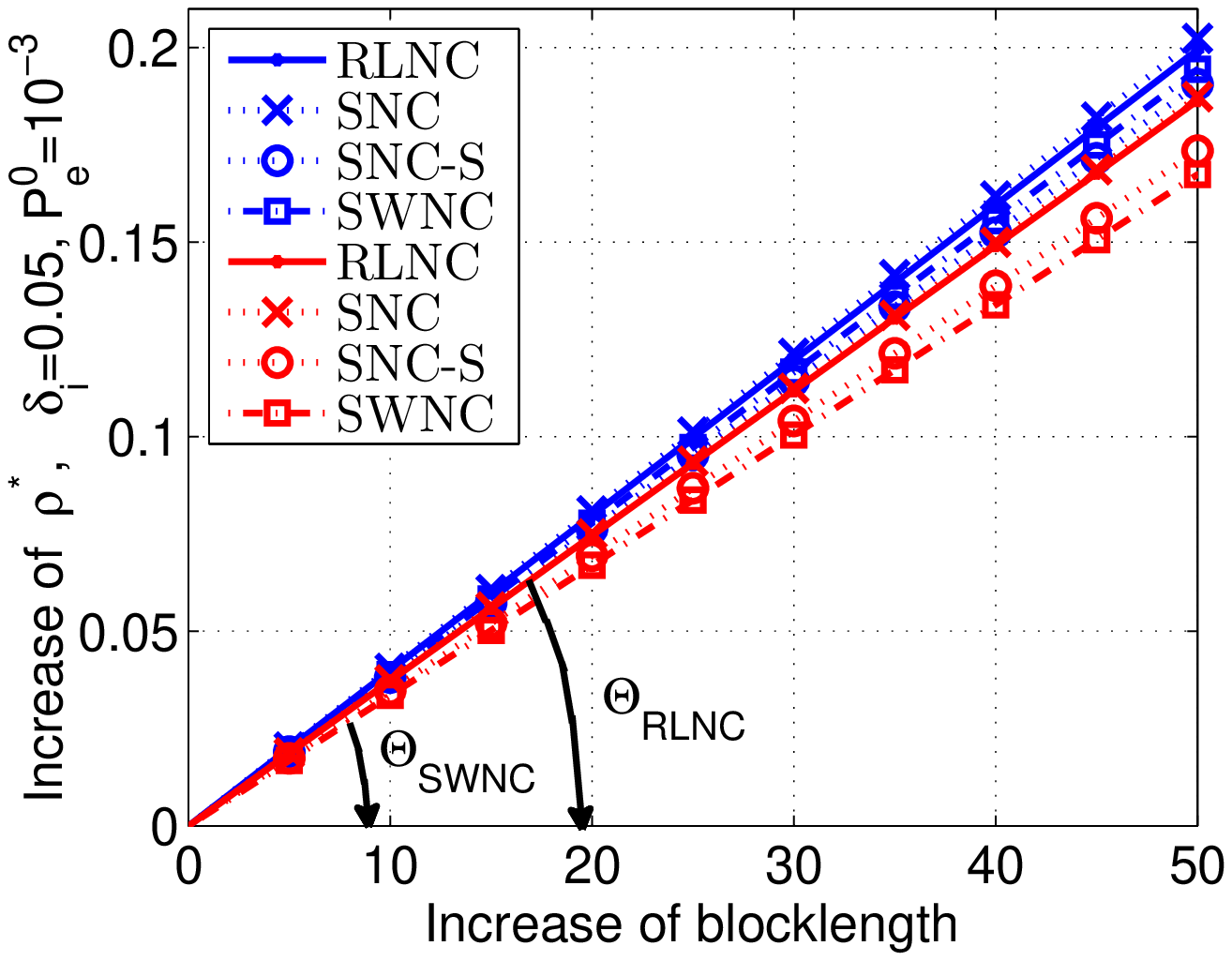,width=0.45\textwidth}}
\hspace{10mm}
\subfloat[$\delta_i=0.2$				\label{fig:SLOPE_E02_Pe10e3_N100}]			{\epsfig{file=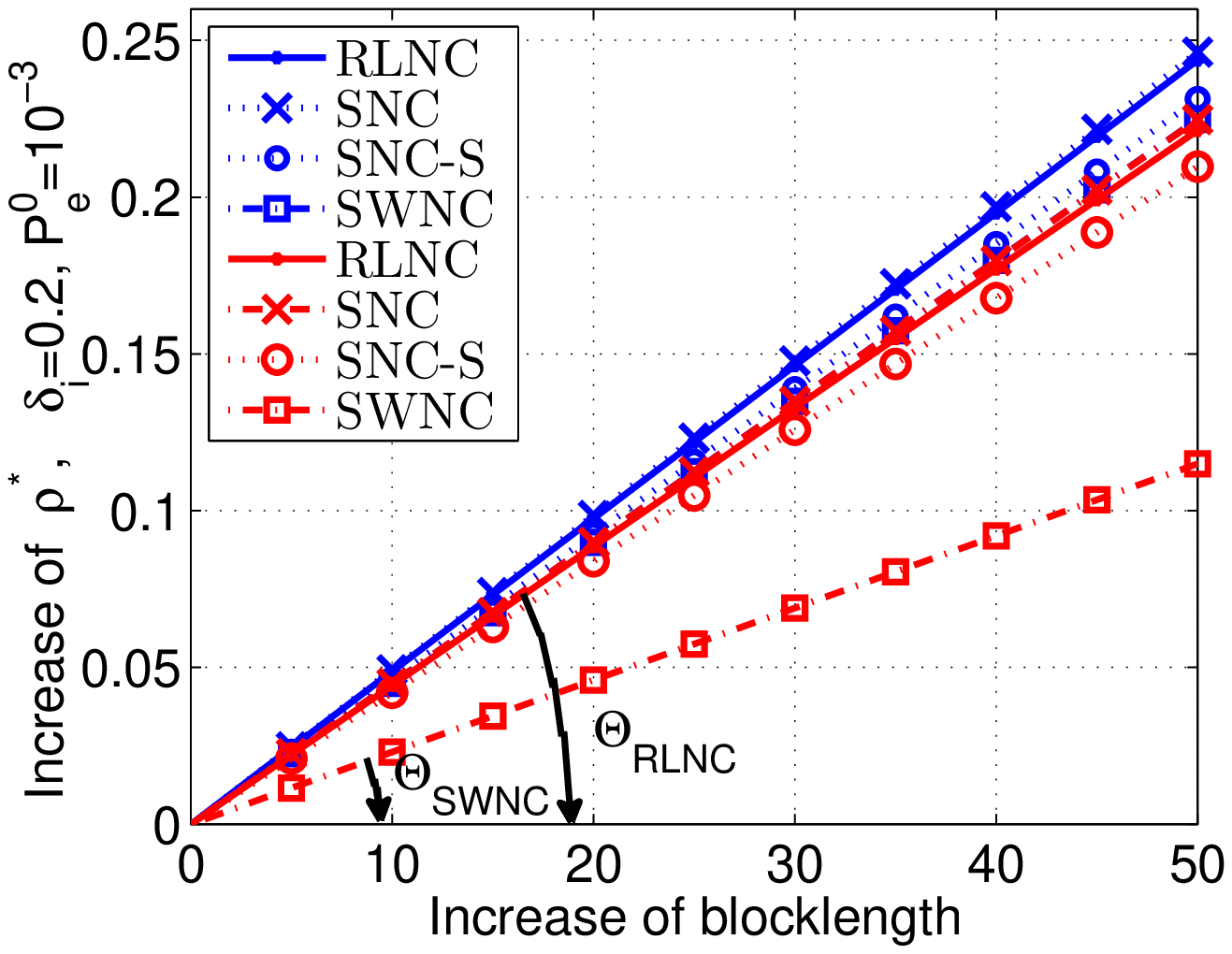,width=0.45\textwidth}}
\par\end{centering}
\caption{The average slope $\Theta$ according to the curves depicted in Fig. \ref{fig:OPTIMAL_CODERATE_Pe0_N100} and the increase of $\rho^{*}$ versus the increase of blocklength, $y=\Theta x$ e.g., $x\in [0,50]$, for $P_e^0=10^{-6}$ (Fig. \ref{fig:SLOPE_E005_Pe10e6_N100}-\ref{fig:SLOPE_E02_Pe10e6_N100}) and $P_e^0=10^{-3}$ (Fig. \ref{fig:SLOPE_E005_Pe10e3_N100}-\ref{fig:SLOPE_E02_Pe10e3_N100}) (blue: $|\mathcal{E}|=2$ hops, red: $|\mathcal{E}|=5$ hops). As illustrated, the angle formed by a line w.r.t. a NC scheme and the $x$-axis denotes the average slope. 
\label{fig:SLOPE_Pe0_N100}}
\end{figure*}

In Fig. \ref{fig:OPTIMAL_CODERATE}, we show the optimal coding rate, $\rho^*$, obtained from Algorithm \ref{Algo:SEARCHING} versus blocklength over 2 and 5-hop line networks for $P_e^0=10^{-6}$. We note that the capacity of the topology is the min-cut of the networks which is given by $\underset{1\le i\le |\mathcal{E}|}{\text{min}} \hspace{0.5mm} \big(1-\delta_i \big)$. Major conclusions can be inferred as follows:

\begin{itemize}
	\item For the same $N$ and $\mathcal{D}$, the larger the number of links $|\mathcal{E}|$, the lower the optimal coding rate \cite{Shrader.2009}. The reason is that as observed from Eq. (\ref{eq:PE_RLNC4}), for any $\rho$, $\prod_{i=1}^{|\mathcal{E}|}1-\phi\Big(sgn(\rho-p_i)\sqrt{2NH(\rho,p_i)}\Big)$ reduces with $|\mathcal{E}|$ $(0\leq\phi(.)\leq 1)$. Hence, $P_e(.)$ increases with $|\mathcal{E}|$.

	\item The figure shows the region of exponential coding rate increase for $N \leq 100$ (highlighted by dashed boxes). Specifically, we depict the region of exponential increase with $P_e^0=10^{-6}$ and $P_e^0=10^{-3}$ in Fig. \ref{fig:OPTIMAL_CODERATE_Pe0_N100}. The curves follow the saturating exponential function $\rho^*(N)=c-ae^{-bN}$ \cite{Fekedulegn.1999}. Hence, the slope as a function of $N$ can be given by $f(N) = (\rho^*(N))^{'} = abe^{-bN}$, which shows that the slope decreases with the blocklength.
	Further, the average slope between the start and end points, $(N_1,\rho^*(N_1))$ and $(N_2,\rho^*(N_2))$, respectively, of a specific curve denotes the average rate of coding rate increase and is given as 	$\Theta=\frac{\rho^*(N_2)-\rho^*(N_1)}{N_2-N_1}=\frac{ae^{-bN_1}-ae^{-bN_2}}{N_2-N_1}.$ 
	
	\item In Fig. \ref{fig:SLOPE_Pe0_N100}, we show the average slope according to the curves depicted in Fig. \ref{fig:OPTIMAL_CODERATE_Pe0_N100}. It shows the intrinsic tradeoff rate-delay since capacity achieving schemes (RLNC non-systematic and systematic) present highest slopes, which decreases with packet scheduling techniques to reduce delay (SNC-S and SWNC). Moreover, the higher the number of links $|\mathcal{E}|$ i.e. the more the re-encoding times, the lower the slope.
	\item As expected, for the same $N$ and $\mathcal{D}$, the lower the $P_e^0$, the lower the $\rho^*$. 
	In addition, for low $P_e^0$ the negative influence of the lost information packets of the immediately preceding group (that have just been removed from the decoding window) on the current decoding process at the SWNC decoder increases with erasure rates, number of links, and number of information packets in each block. Therefore, when $N$ is sufficiently large, the optimal coding rate for SWNC is lower than that of SNC-S.
\end{itemize}

%%------------------------------------------------------------------------------------------------------------------------------------------
\section{Conclusions \label{Sec:CONCLUSIONS}}
In this paper, we proposed a methodology to compute the optimal finite-length coding rate for random linear NC schemes over a line network. We apply our method for representative erasure rates, target PLRs and number of hops. We model the exponential increase with $N$, which reveals the difference between capacity-achieving and non-achieving schemes.

%%------------------------------------------------------------------------------------------------------------------------------------------

% Bibliography file
\bibliographystyle{IEEEtran}
\bibliography{RELIABILITY}

\end{document}